# Analysis of Hydrogen Production Costs across the United States and over the next 30 years

*Mahmoud M. Ramadan[1*], Yuanchen Wang[2], Pragya Tooteja[3]*

**Abstract**

Hydrogen can play an important role for decarbonization. While hydrogen is usually produced through SMR, it can also be produced through water electrolysis which is 'cleaner'. The relative cost and carbon intensity of hydrogen production through SMR and electrolysis vary throughout the United States because of differences in the grid. While many hydrogen cost models exist, no regional hydrogen study has been conducted across the US. We studied how the Levelized Cost of Hydrogen (LCOH) and carbon intensity for producing hydrogen vary across the US. We looked at electrolysis technologies (Alkaline, PEM, and SOEC) and compared them to SMR. In 2020, SMR with 90% CCUS has a lower average LCOH and carbon intensity for hydrogen production than electrolysis by SOEC. For states with cleaner grids, hydrogen produced through SOEC has a lower carbon intensity than hydrogen produced using SMR with 90% CCUS. Washington has one of the lowest carbon footprints and the lowest LCOH to produce hydrogen through electrolysis (alkaline). We predict that the LCOH for hydrogen production will be $3.2/kg for Alkaline, $3.1/kg for PEM, and $2.6/kg for SOEC by 2050 with constant electricity prices. These projected LCOHs are still higher than the LCOH for hydrogen produced through SMR with 90% CCUS. If electricity costs decrease to 2c/kWh, we expect to reach cost-parity with SMR with 90% CCUS. The results suggest that significant investment in decarbonizing the grid and lowering the cost of electricity needs to be made to make electrolysis more competitive compared to SMR.

## Background and Introduction

Hydrogen has a critical role in addressing critical climate and environmental goals. The methods used to produce hydrogen make it less 'green' than originally thought. In the US, the majority of hydrogen is produced from natural gas through steam-methane reforming (SMR)[1].

Producing hydrogen through SMR releases carbon dioxide into the atmosphere. The main reaction to produce hydrogen through SMR (called the steam-methane reforming reaction) is as follows:

$$CH_4 + H_2O \text{ (+ heat)} \rightarrow CO + 3H_2$$

This reaction is followed by a water-gas shift reaction to produce further hydrogen and is as follows:

$$CO + H_2O \rightarrow CO_2 + H_2 \text{ (+ small amount of heat)}$$

Although carbon monoxide and carbon dioxide are produced through SMR, a large percentage of the carbon emissions can be mitigated through carbon capture and sequestration (CCS).

[1]Department of Materials Science and Engineering, Massachusetts Institute of Technology, Cambridge, MA
[2]Harvard Kennedy School, Harvard University, Cambridge MA
[3]Harvard Business School, Harvard University, Cambridge MA
*Email: ramadanm@mit.edu

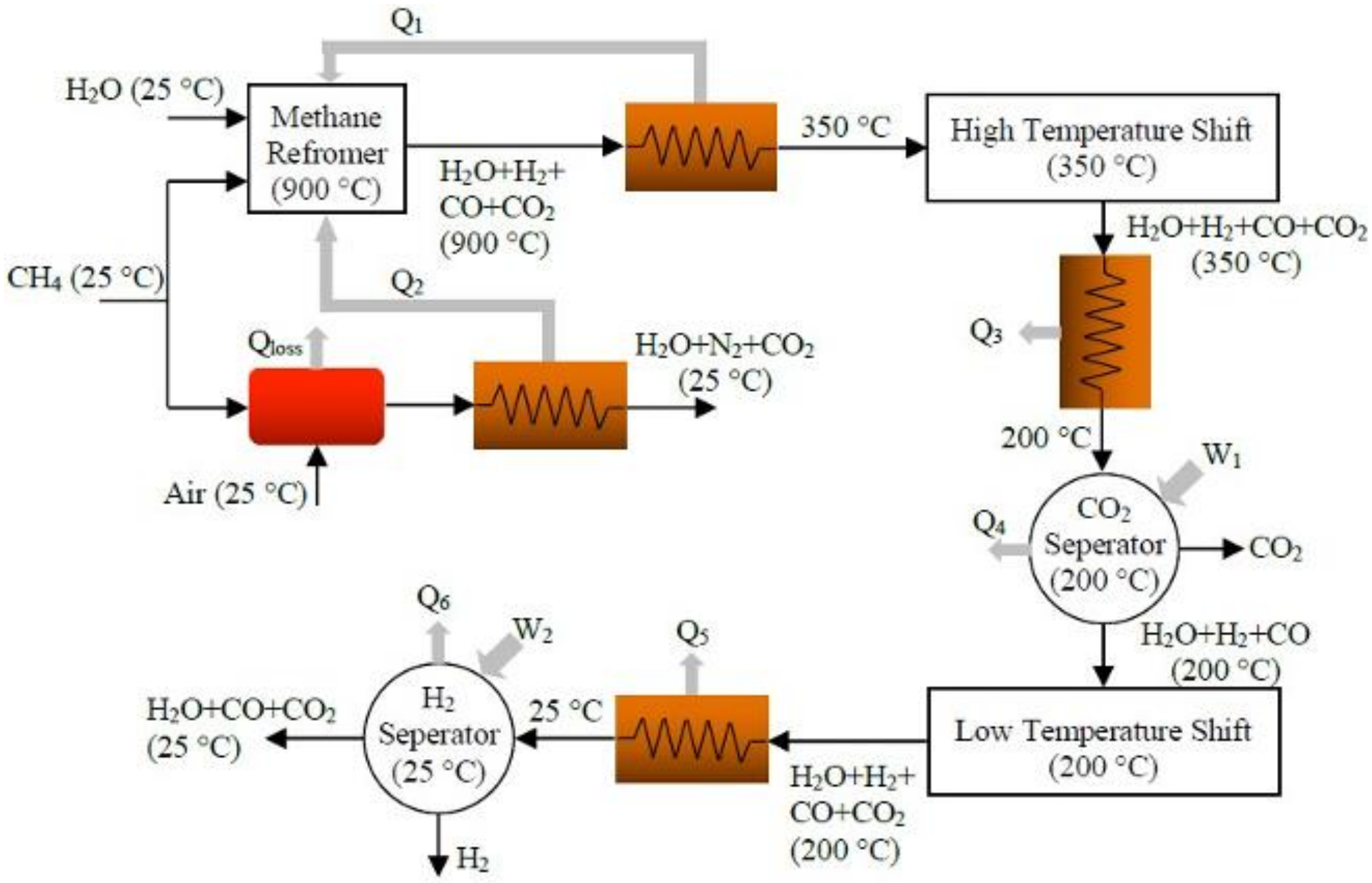

**Figure 1.** Process flow diagram of SMR. Adapted from MIT Course 2.60 HW3 (2022).

Electrolysis is an attractive alternative because it is a "green" process and produces hydrogen by splitting water electrochemically to produce hydrogen. However, the process is not carbon-free: the hydrogen produced through electrolysis is only as green as the electricity used in the process itself

$$2H_2O \rightarrow O_2 + 2H_2 \qquad \Delta H = +286 kJ/mol\ H_2$$

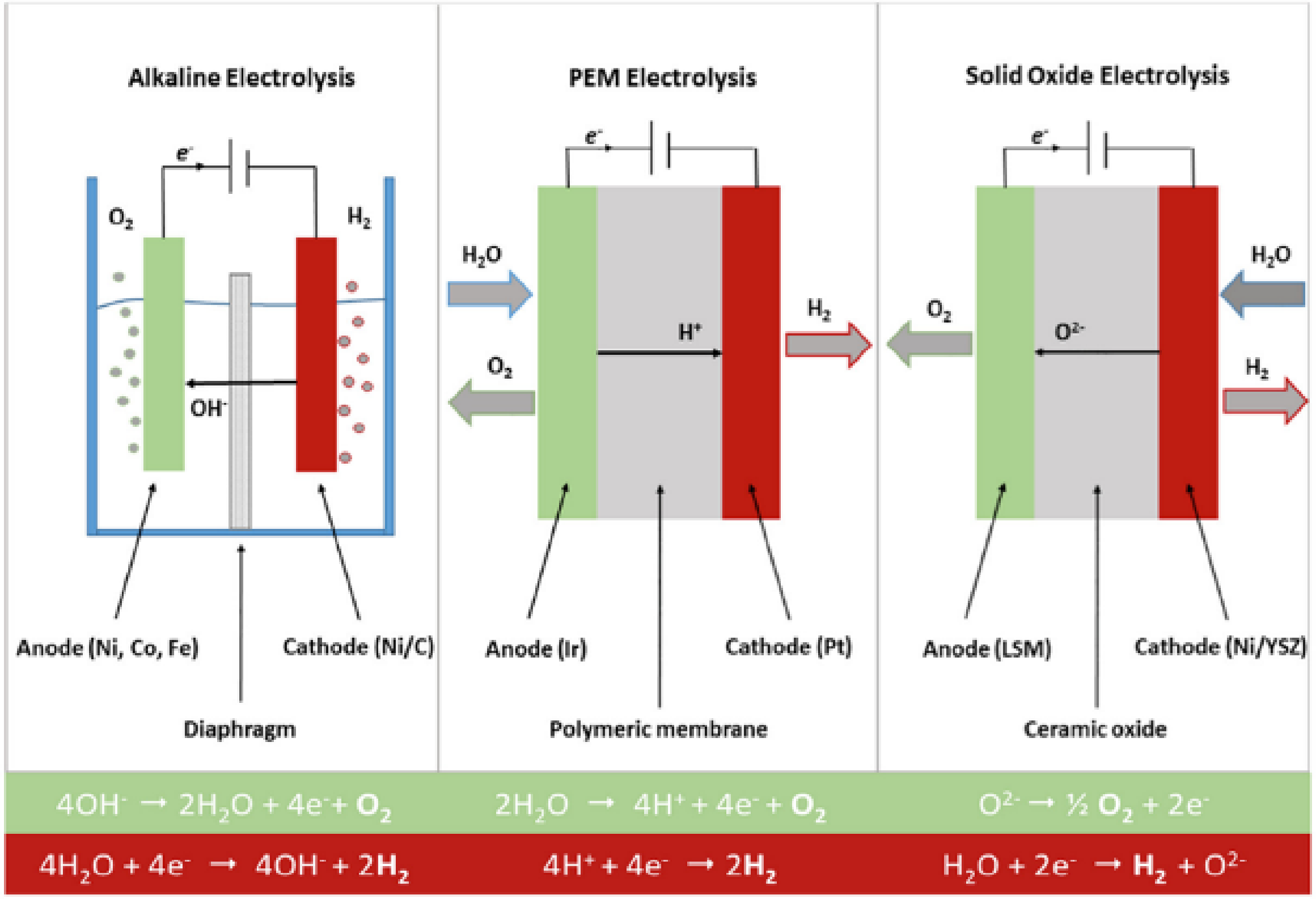

**Figure 2.** Alkaline electrolyzers, polymer membrane electrolyzers, and solid oxide electrolyzers (from left to right). Despite fundamental differences in chemistry, ion transport, and electrolyte material, it can be seen here that the general setup is very similar across all three. Also of note is that these are not general descriptions. The solid oxide electrolyte shown here uses oxygen ions as charge carriers, for example, while others might use proton-conducting electrolytes where $H^+$ ions are transferred[2].

We studied three electrolysis technologies in the paper: Alkaline, Polymer Electrolyte Membrane (PEM) and Solid Oxide Electrolysis Cell (SOEC). The electrolyte of alkaline electrolysis is Potassium Hydroxide (KOH) at 5-7 mol/L. PEM and SOEC adopt solid electrolyte, which usually are Perfluoroacidsulfonic (PFSA) membranes and Yttria-stabilizedZirconia (YSZ) respectively[3]. In 2020, 61% of global installed electrolysis capacity is Alkaline and 31% is PEM[4]. Alkaline became commercially available starting in the 1920s. Currently, alkaline has the lowest capital cost among all the three technologies. However, it requires a large area and has relatively low efficiency. PEM requires a smaller area. However, the electrolysis cell includes precious materials such as Iridium (Ir), Platinum (Pt) and Gold (Au). Currently, PEM ($700 -$1,400/kW) has a higher capital cost than Alkaline ($500-$1,000/kW). The lifespan and reliability of large-scale PEM projects is yet to be validated. SOEC operates in 550-1,000 degree Celsius and has the highest efficiency. However, with the highest cost ($2,500/kW) and the least flexibility, the technology is still at pilot stage. Although performance ranges widely by technology today, these gaps are expected to close over time. There is no clear winner across all applications. Each technology has its own challenges, from critical materials to performance, durability and maturity. There is active R&D on all three technologies to drive down the costs and compete with each other.

The purpose of this paper is to compare the present costs and carbon-intensity of producing hydrogen through SMR and through three different electrolysis technologies (Alkaline, PEM, and SOEC) on a regional level within the US. Our analysis can enable policy makers to make more informed decisions when deciding how to roll-out hydrogen production in the US.

Much literature has studied the cost and carbon intensity of electrolysis which is summarized in the following table. However, they neither disclose the calculation method and data source nor distinguish different electrolysis technologies. There is also no comprehensive calculation for the electrolysis costs and carbon intensity in the United States by states with most recent data.

**Table 1.** Literature Review

| **Name** | **Content** | **Gaps/Comments** |
|---|---|---|
| Goldman Sachs Carbonomics: The Clean Hydrogen Revolution[5] | Market size, competitiveness cost for different applications, LCOH, cost reduction trajectory, electrolysis installed capacity, amount of upfront investment | No carbon intensity. Did not explicit electrolysis technology. Did not disclose data and methods. No detailed study of the U.S. by states. |
| Bernstein Hydrogen Highway 2020: Ready for Prime Time | Cost comparison among FCEV and BEV and ICE, market size, LCOH evolution | No carbon intensity. Did not explicit electrolysis technology. Did not disclose data and methods. No detailed study of the U.S. by states. Last updated in 2009. No distinction among the three electrolysis technologies. |
| Bernstein Hydrogen Highway 2021: Hydrogen One, Carbon Zero | Market size, competitiveness cost for different applications, LCOH, cost reduction trajectory, cost-competitiveness with different applications | |
| Hydrogen Council Path to Hydrogen Competitiveness: A Cost Perspective[6] | Learning rate, breakeven cost, cost trajectory, LOCH with varying load factors, cost-competitiveness with different applications | |
| IRENA Green Hydrogen Cost Reduction [7] | Capital costs, learning rate, efficiency | |
| IEA Global Hydrogen Review 2021[4] | Demand of hydrogen under policy targets of different countries | |
| NREL Hydrogen Demand and Resource Analysis[8] | Hydrogen demand and production cost by SMR and electrolysis | |
| NREL Hydrogen Analysis | Costing and financial parameters | Capital costs last determined in |
| (H2A) Production Models[9] | for SMR plants | 2018 |
| U.S. Energy Information Administration (EIA)[10] | Price of industrial electricity and natural gas by state | Latest full set of data is from Year 2020 |

## Methods

### Electrolysis

The levelized cost of hydrogen (LCOH) for the three technologies are calculated as follows. The data sources are shown in Figure 3.

**(1)** $$Present\ Value\ Interest\ Factor\ of\ Annuity\ (PVIFA) = \frac{1-(1+Discount\ Rate)^{-Lifetime}}{Discount\ Rate}$$

**(2)** $$Capital\ Cost = Unit\ System\ Cost * Capacity$$

**(3)** $$O\&M\ Cost = Unit\ O\&M\ Cost * PVIFA$$

**(4)** $$Electricity\ Cost = Unit\ Electricity\ Cost * Capacity * 1\ year * PVIFA$$

**(5)** $$Hydrogen\ Production = \frac{Capacity * 1\ year * PVIFA}{Efficiency}$$

**(6)** $$\boldsymbol{LCOH = \frac{Capital\ Cost + O\&M\ Cost + Electricity\ Cost}{Hydrogen\ Production}}$$

**(7)** $$\boldsymbol{Unit\ Capital\ Cost\ 2050 = Unit\ Capital\ Cost\ 2020 * (1 - Learning\ Rate) * \log_2(\frac{Cumulative\ Production\ 2050}{Cumulative\ Production\ 2020})}$$

**(8)** $$\boldsymbol{H_2\ Carbon\ Intensity = Grid\ Carbon\ Intensity * Efficiency}$$

| | | Alklaine | PEM | SOEC | Source: |
|---|---|---|---|---|---|
| Learning Rate (APS) | % | 14.5% | 14.0% | 10.5% | IRENA |
| Learning Rate (NZE) | % | 14.0% | 13.5% | 10.0% | IRENA |
| 2020 Cumulative Production | MW | 20,000 | 90 | 2 | IEA, IRENA |
| | | | | | |
| Capacity | kW | 10,000 | 10,000 | 1,000 | IRENA, company data |
| Lifetime | 1,000 hr | 60 | 75 | 40 | IRENA, company data |
| Efficiency | kWh/kgH2 | 56 | 51 | 44 | IRENA, company data |
| Unit System Cost | USD/kW | 750 | 1,200 | 2,500 | IRENA, company data, DOE |
| Unit Electricity Cost | USD/kWh | 0.1 | 0.1 | 0.1 | EIA |
| Unit O&M Cost | USD/yr | 1,800 | 1,500 | 20,000 | Company data |
| Discount rate | % | 7% | 7% | 7% | US Treasury Bond |
| Present value interest factor | % | 5 | 6 | 4 | |
| **LCOH** | **USD/kg H2** | **7** | **6** | **8** | |
| | | | | | |
| Grid Carbon Intensity | kg CO2 equavalent/kWh | 0.2 | 0.2 | 0.2 | EPA |
| **H2 Carbon Intensity** | **kg CO2 equavlent/kg H2** | **11** | **10** | **9** | |

**Figure 3**. Data sources for LCOH and carbon intensity of electrolysis. Data in blue are inputs and data in black are outputs. APS indicates Announced Pledges Scenario and NZE indicates Net Zero Emissions Scenario of IEA estimates. The unit electricity cost data and carbon intensity data are placeholders for variation by states.

**Life Cycle Assessment of Steam-Methane Reformation**

A widely relied-upon reference in the literature is the 2001 National Renewable Energy Laboratory (NREL) Life Cycle Assessment (LCA) of Hydrogen Production via Natural Gas Steam Reforming[11].A cradle-to-grave LCA on an SMR plant determined that 99% of air emissions by mass is $CO_2$, while $CO_2$ accounts for 89.3% of the global warming potential over 100 years (GWP100) of the system, with $CH_4$ and $N_2O$ accounting for the rest.

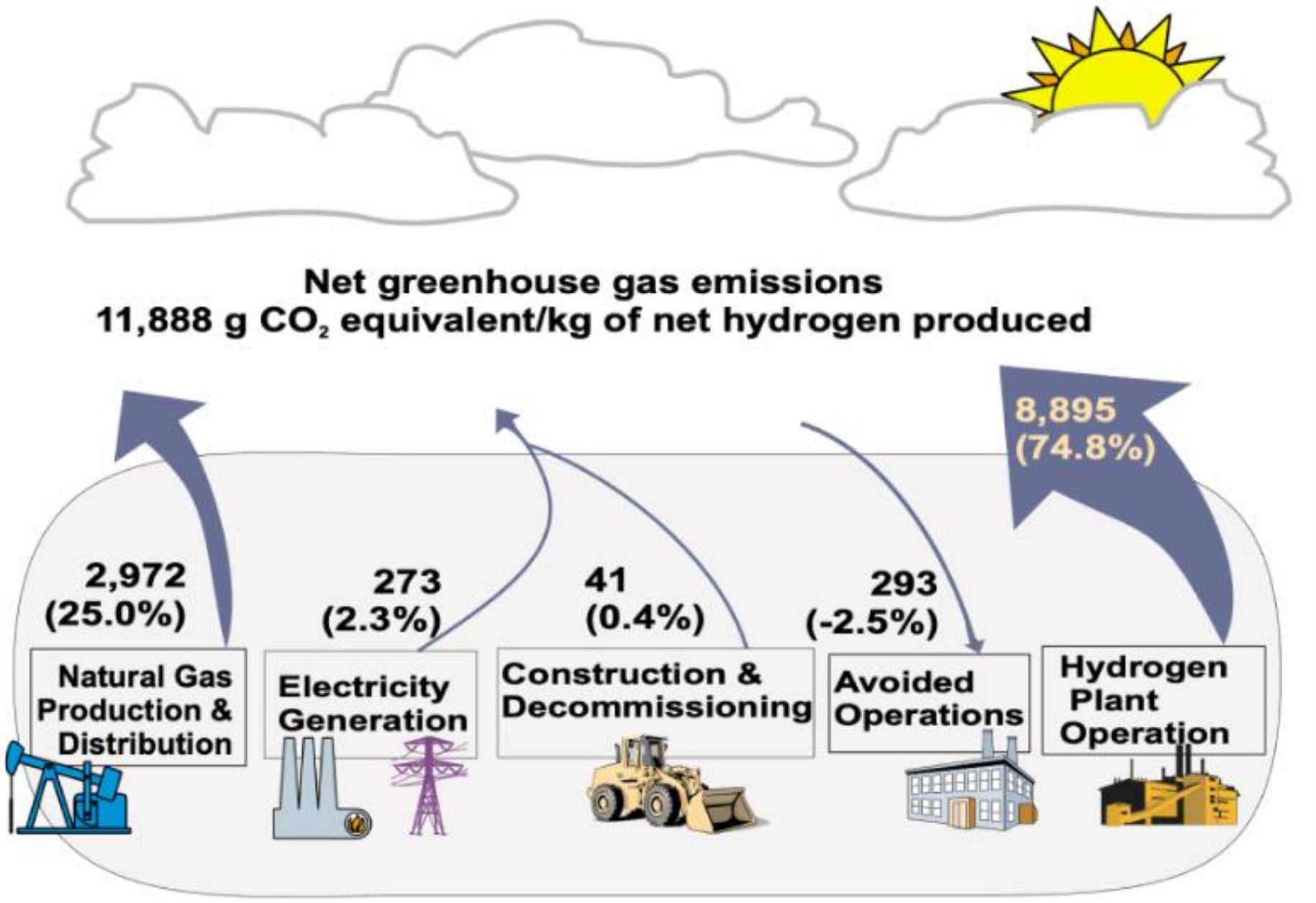

**Figure 4.** Sources of System Global Warming Potential [11]. (a) Construction and decommissioning include plant construction and decommissioning as well as construction of the natural gas pipeline. (b) Avoided operations are those that do not occur because excess steam is exported to another facility. See section 3.0 for more information about this.

Recent studies have determined that EPA under-reports methane leakage in the U.S.[12], [13]. A 2022 study co-published by the MIT Energy Initiative re-conducted a LCA on SMR with different methane leakage rates[14]. For our model, we chose a methane leakage rate of 3.0% is assumed throughout the entire U.S. By linearly interpolating across the data in the published paper[14], the SMR carbon footprint is determined to be 12.9 kg $CO_2$-equivalent/ kg-$H_2$ without CCS and 5.3 kg $CO_2$-equivalent /kg-$H_2$ with 90% CCS respectively (*Note that only $CO_2$ is captured and not methane).

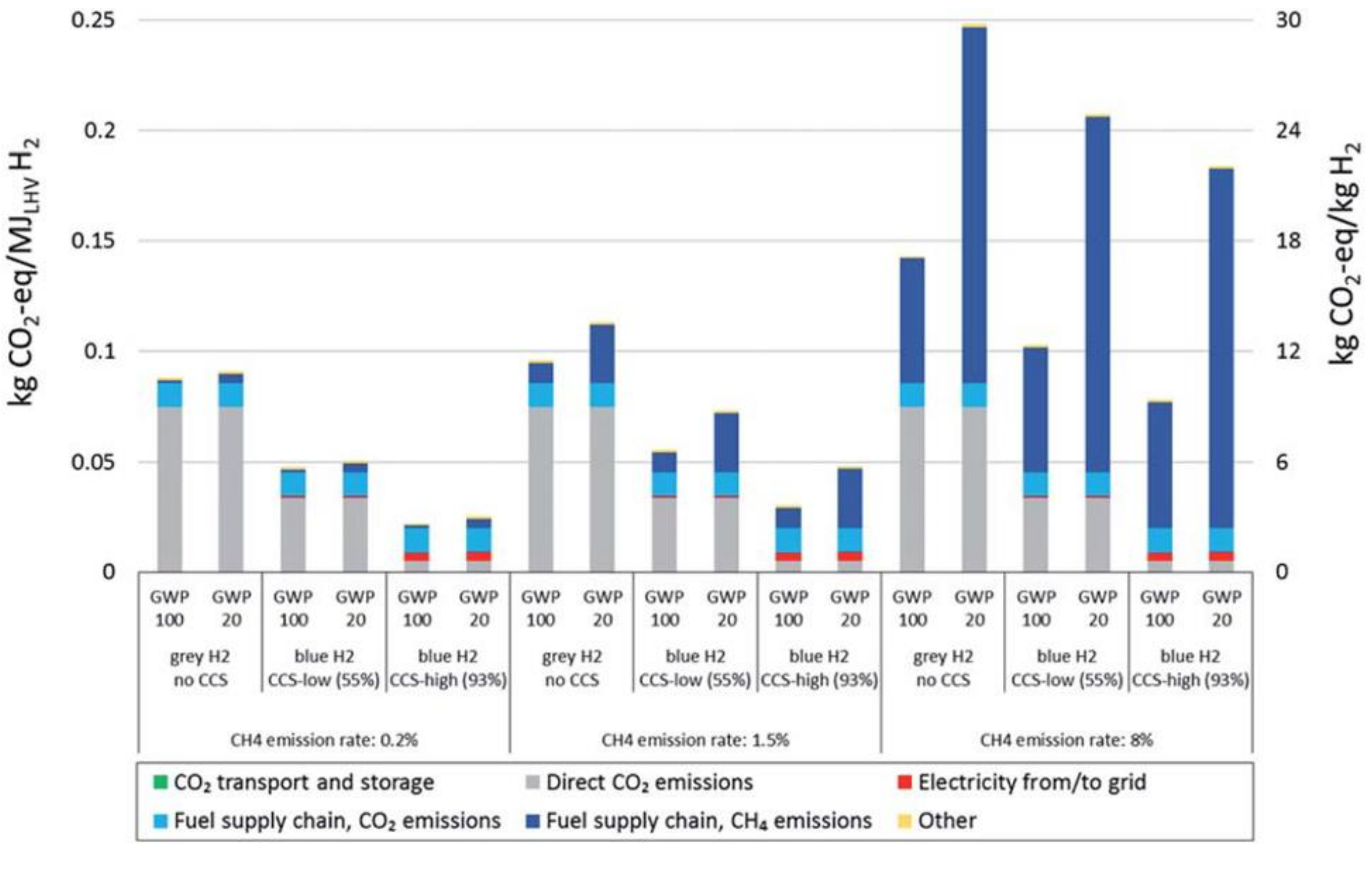

**Figure 5**. Impacts on climate change associated with the production of NG-based hydrogen with methane emission rates of 0.2%, 1.5%, and 8%, and two plant configurations with high and low CO2 removal rates [14].

### Cost of SMR

The NREL Hydrogen Analysis (H2A) Production Models are technology-specific spreadsheet files that contain macros necessary to compute a hydrogen price calculation[9]. The H2A Production Models were utilized to obtain cost estimates of hydrogen produced by SMR, both with and without CCS. The natural gas price is a major contributor to the price of hydrogen produced during SMR.

The relevant calculations for the SMR plant (materials and energy balances) in the H2A Production Models were conducted in Aspen, and the equipment costing is based on grass-roots cost estimates and commercial offering as documented by the NREL H2A team. In the model, the SMR plant, through the pressure swing adsorption (PSA) process produces hydrogen in a stream at 99.6% purity. In the case where CCS is considered, an amine gas treatment process is modeled to capture 90% of the $CO_2$ emitted by adding the relevant equipment including an absorber column and a stripper column.

The NREL SMR production model includes the capital (machinery) and operating costs, and all the relevant financial parameters[9]. The capital and financial parameters were kept constant throughout the analysis. In order to compare the production cost of hydrogen using SMR, the only two parameters in the model that were varied throughout the analysis were the cost of electricity and the cost of natural gas. The industrial cost of electricity and industrial cost of natural gas for each state were used from the EIA database, in order to compare hydrogen production costs throughout different states. The latest complete set of state costs were from the year 2020 (2021 data are not complete yet for all states); therefore, 2020 costs were used throughout the analysis.

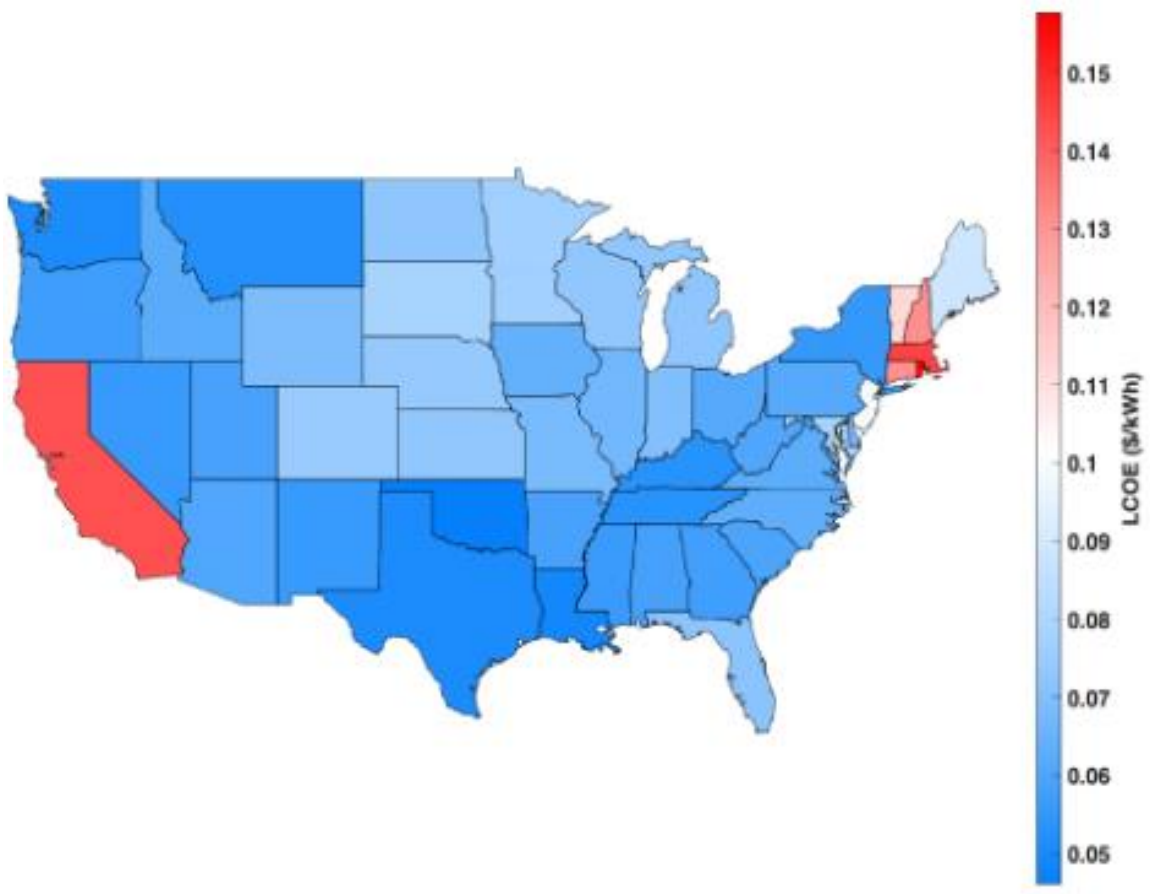

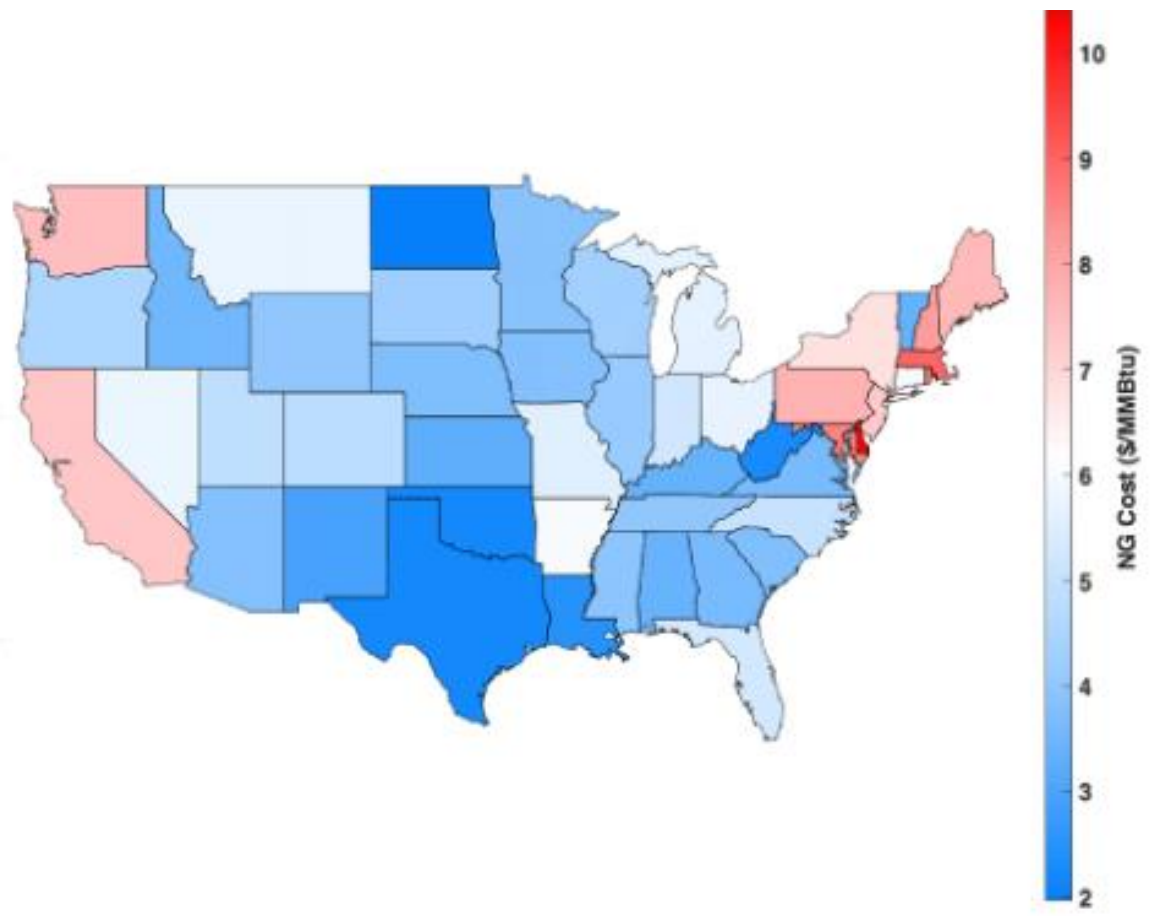

**Figure 6**. Levelized cost of industrial electricity in USD per kWh by state (left graph) and cost of industrial natural gas in USD per metric Million British Thermal Units (MMBtu) by state (right graph), as reported by EIA for the Year 2020[15], [16].

## Results

### Electrolysis

In 2020, the average cost of LCOH is $4.6/kg for Alkaline, $4.5/kg for PEM, and $6.3/kg for SOEC. The results align with DOE calculation. Electricity costs account for over 50% of total LCOH. The most economical states ($3.5-3.7/kg H2) to produce hydrogen using electrolysis are the states with the lowest industrial electricity price, such as Oklahoma, Louisiana, Texas and Washington. The most expensive states (~$8-14/kg H2) to produce hydrogen with electrolysis are Hawaii, Alaska, Rhode Island, Massachusetts, and California.

In 2050, if we give all the three technologies the same production capacity to drive down the cost curve. SOEC will become the cheapest technology due to its high efficiency. If we assume the same electricity cost as 2020. In 2050, the average cost of LCOH is $3.2/kg for Alkaline, $3.1/kg for PEM, and $2.6/kg for SOEC. Electricity costs account for almost 100% of LCOH since the extended lifespan, declined capital cost, increased reliability make capital costs and O&M costs negligible. The differences among states remain the same ranking as 2020 but the disparity further increases since electricity cost plays a dominant part in LCOH.

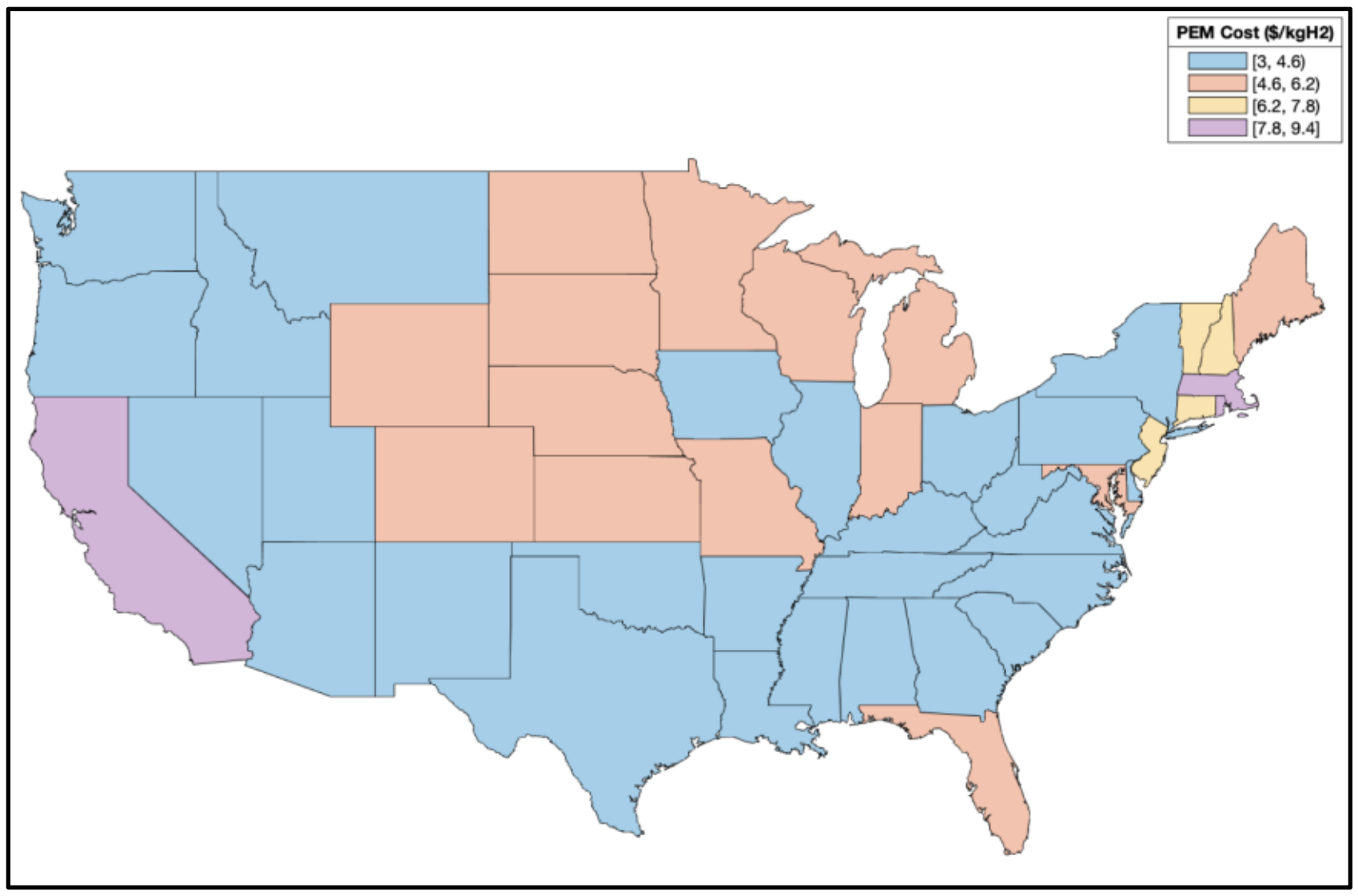

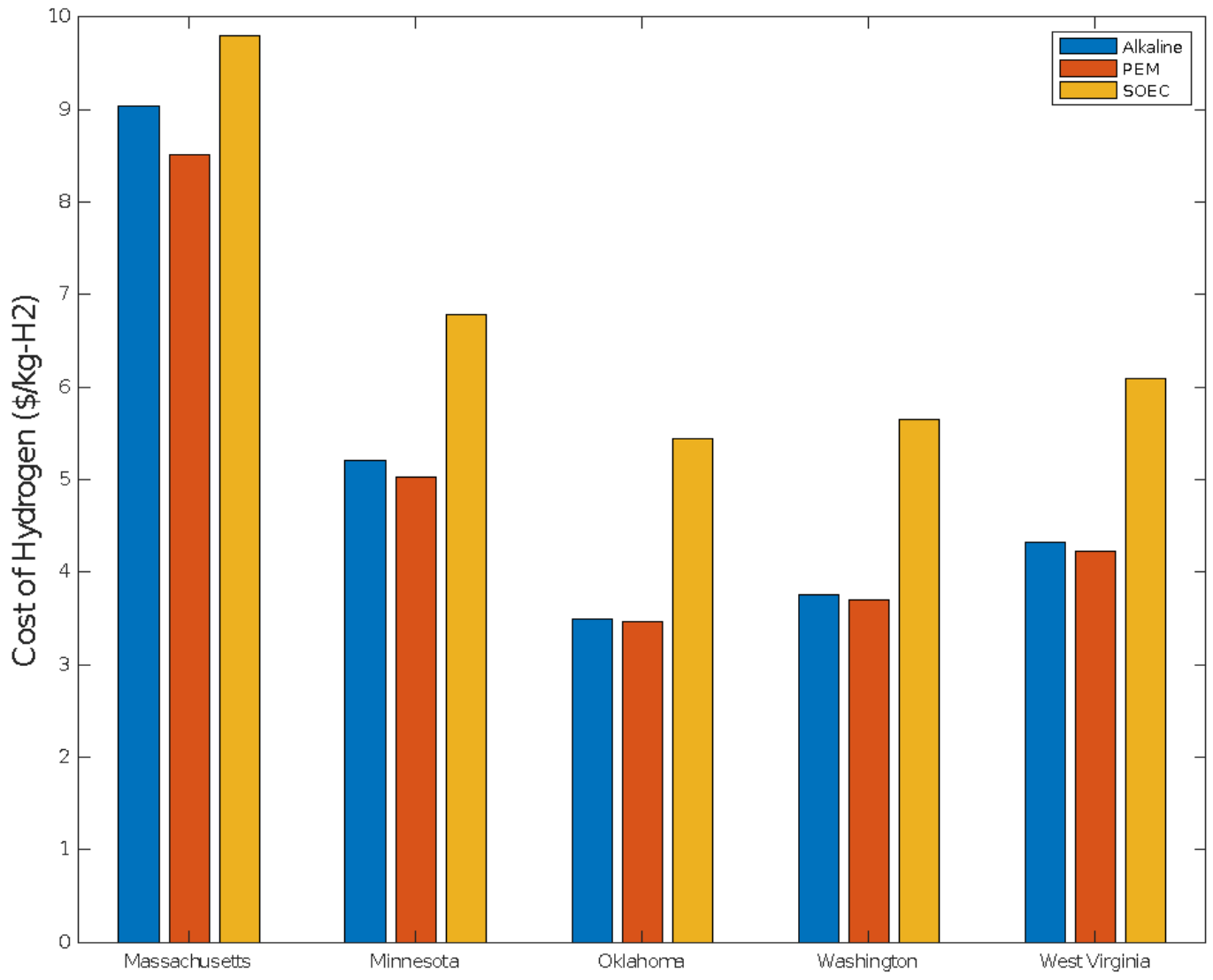

**Figure 7**. Levelized cost of Hydrogen produced by PEM in USD per kg hydrogen by state for the Year 2020. The 2050 LCOH and the LCOH of the other two technologies follow the same pattern.

In 2020, the average carbon intensity is 21kg CO2 equivalent/kg H2 for Alkaline, 19kg CO2 equivalent/kg H2 for PEM, and 16kg CO2 equivalent/kg H2 for SOEC. The results align with the Hydrogen Council calculation. Since we only calculate the carbon intensity of electricity for hydrogen production, the higher the efficiency, the lower the carbon intensity. The disparity among states could be as high as 50x difference due to the different grid carbon intensity. The cleanest states (1-5kg CO2 equivalent/kg H2) are Vermont, Washington, Idaho, Maine, and New Hampshire. The dirtiest states (43-50kg CO2 equivalent/kg H2) are Wyoming, West Virginia, and Kentucky. Starting in 2035, the carbon intensity of electrolysis will approach zero according to the 2035 zero-grid proposal of the Biden Administration.

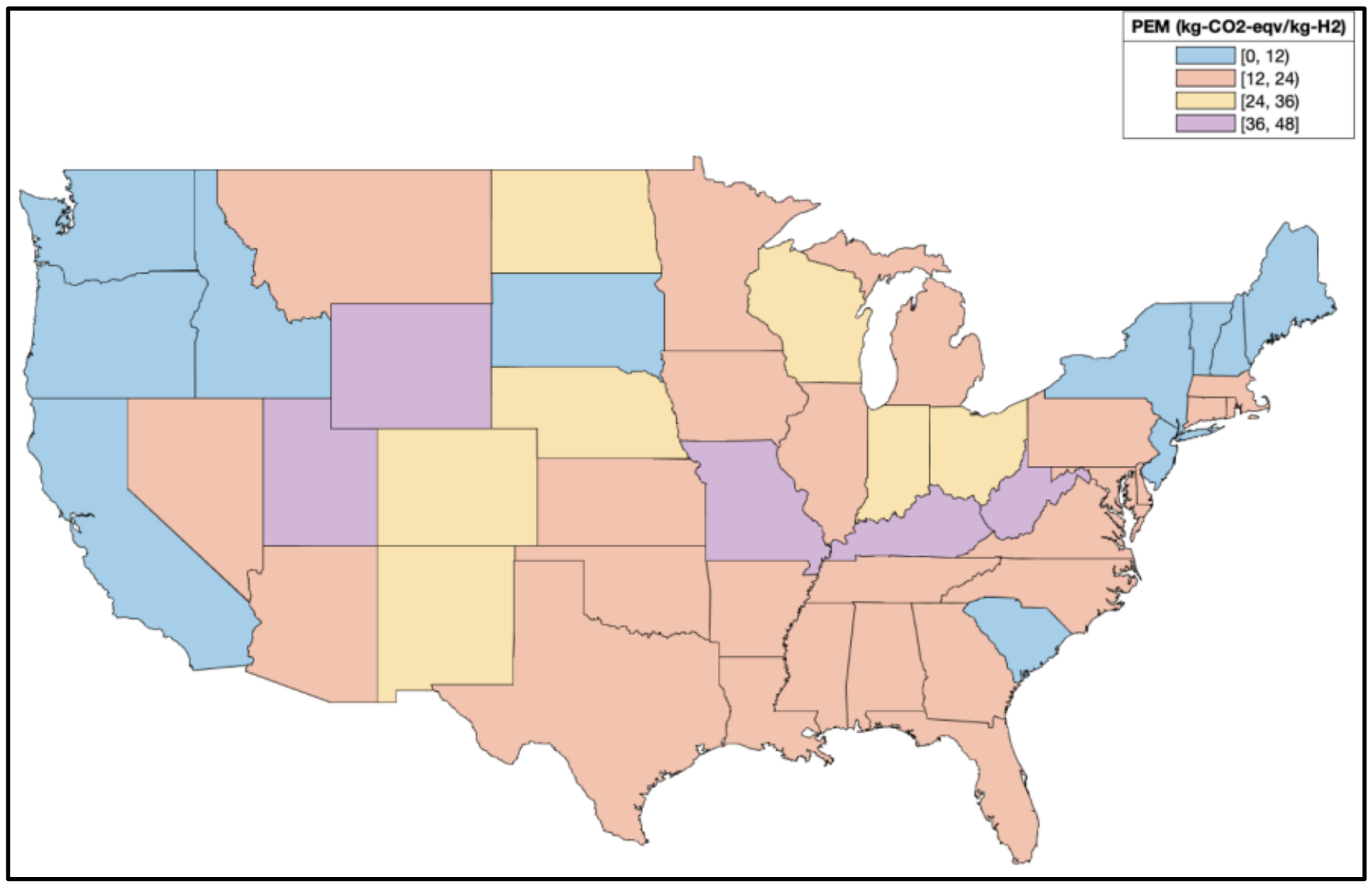

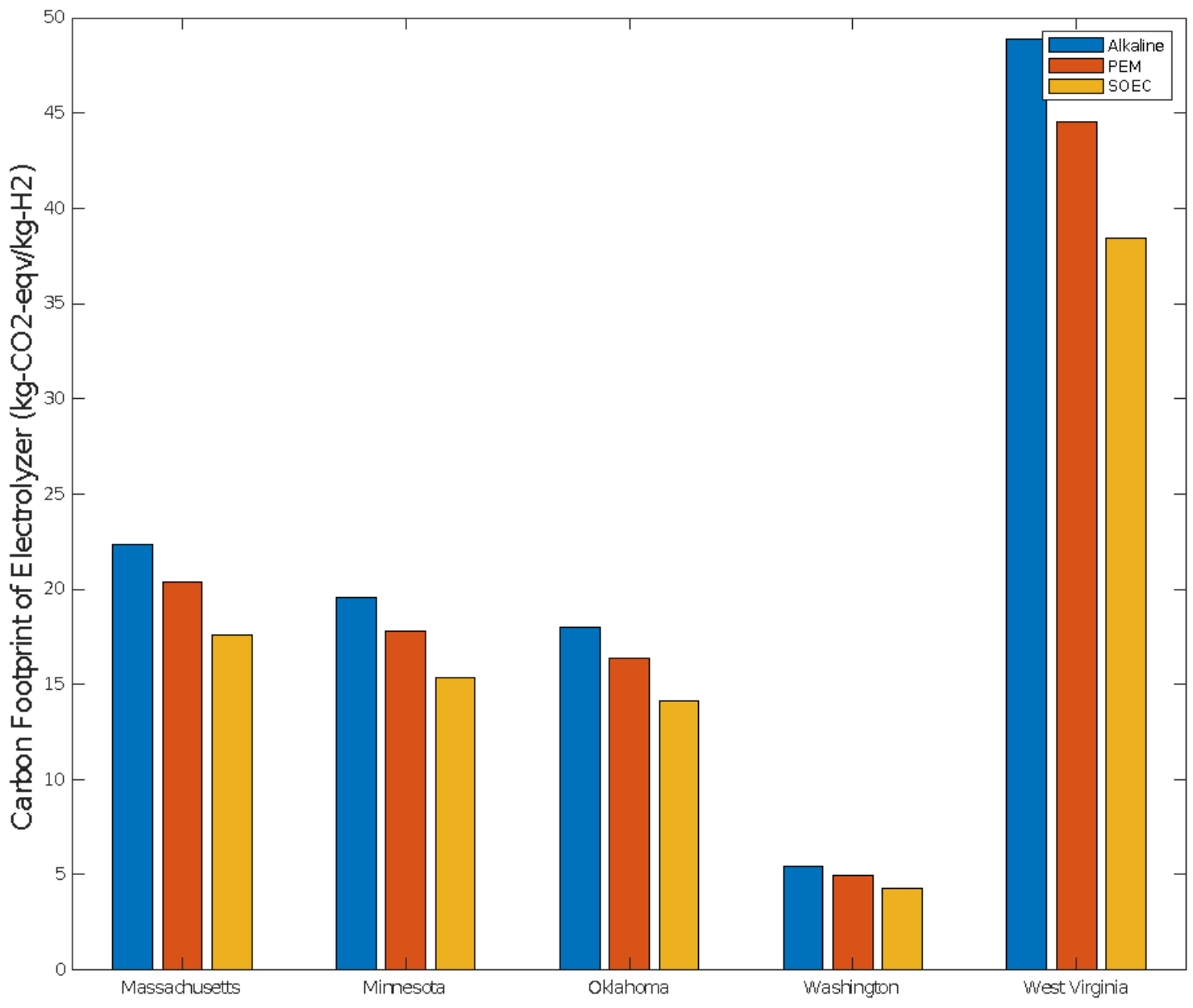

**Figure 8**. Carbon Intensity of Hydrogen produced by PEM in kg CO2 equivalent per kg hydrogen by state for the Year 2020. The carbon intensity of the other two technologies follow the same pattern.

Considering both carbon intensity and costs, Washington and Idaho are the most optimal states to produce low cost hydrogen with a low carbon footprint in 2020. Massachusetts and Rhode Island are the worst states for electrolysis with both high costs and carbon footprints.

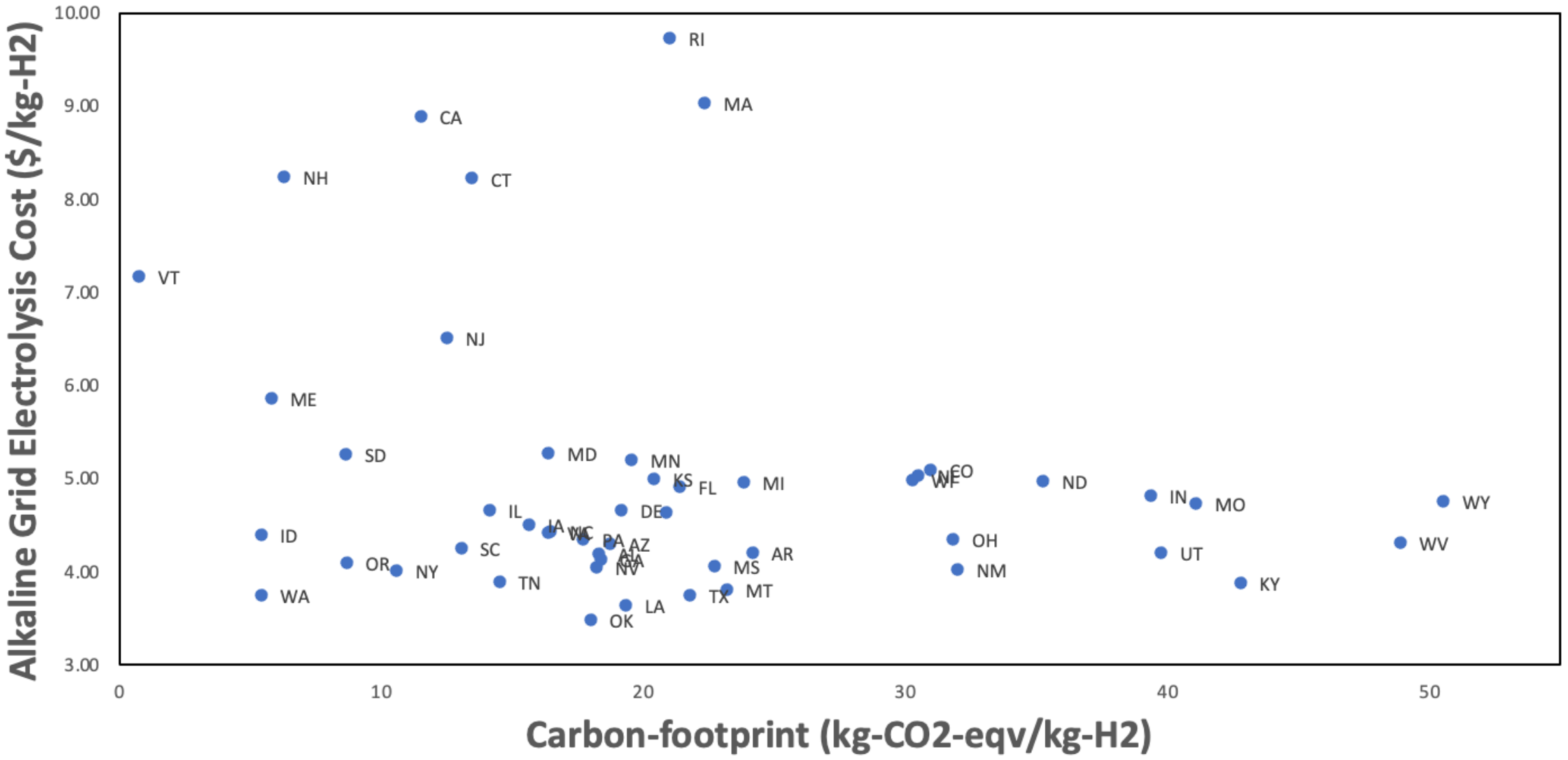

**Figure 9**. Comparison of Cost and Carbon intensity of electrolysis in different states.

**SMR**

The cost of hydrogen produced via SMR with and without CCS was calculated using the NREL H2A production models. The cost of SMR (without CCS) for each state is shown in color map format in Figure 10. As shown in Figures 7 and 10, the cost of SMR is a strong function of the cost of natural gas and is not very dependent on the cost of electricity.

The specific cost of hydrogen produced in each state is shown in Figure 11. The extra cost of CCS added to the SMR process is $0.4/kg-$H_2$.

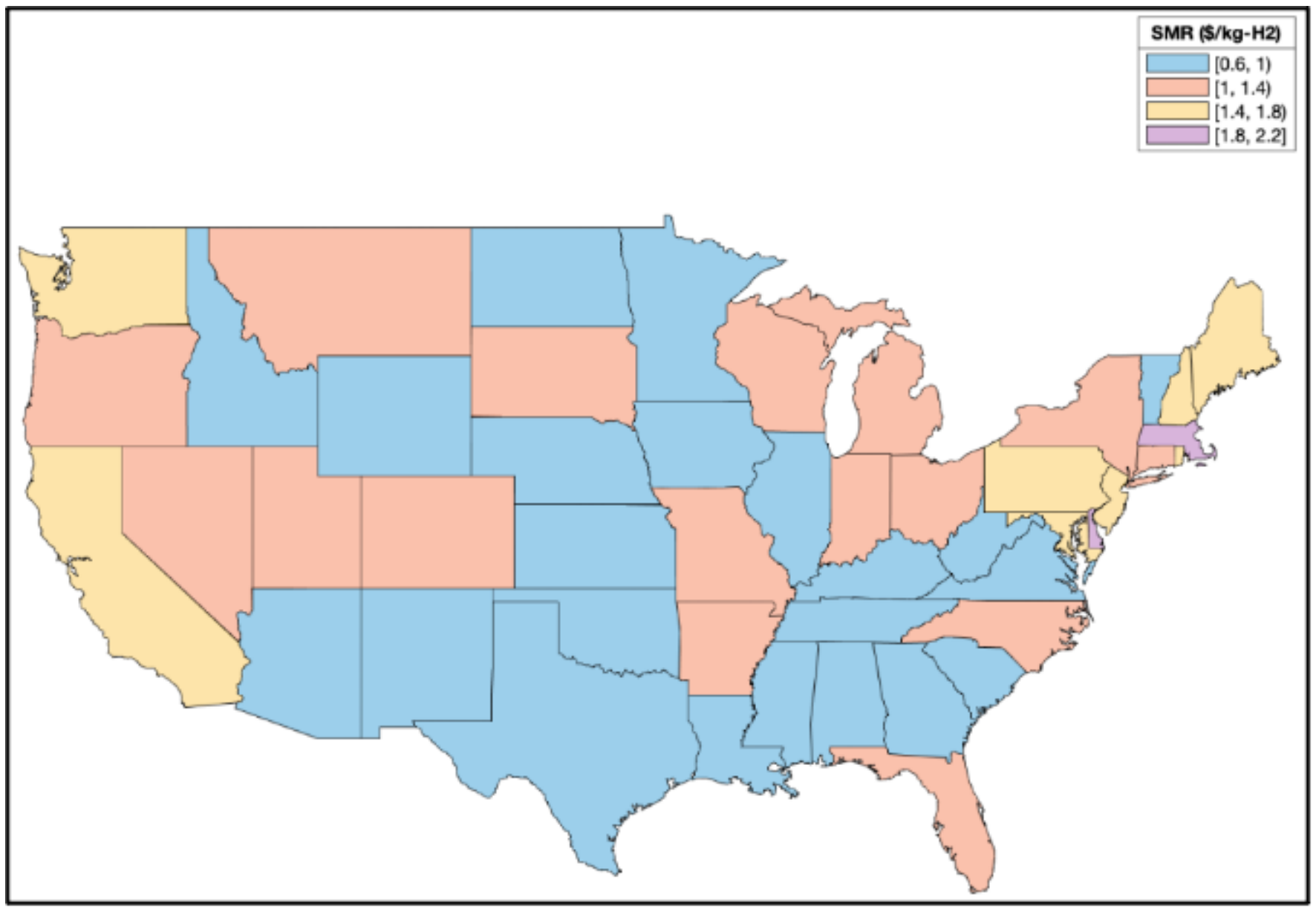

**Figure 10.** Color map of the U.S. showing our calculated results of the cost of SMR without CCS in $/kg-$H_2$ for each state.

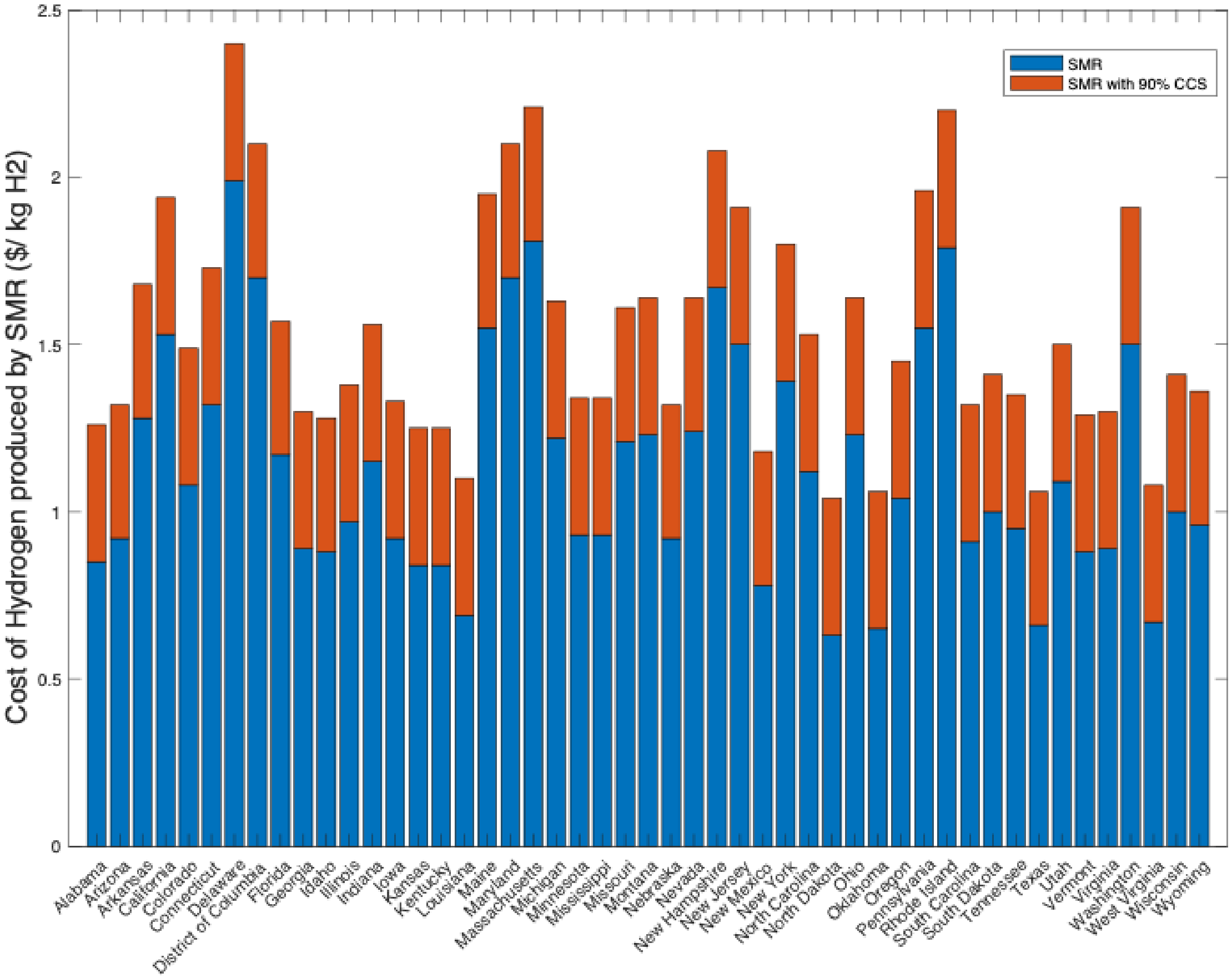

**Figure 11.** Tabular form of calculated SMR cost with and without CCS in $/kg-$H_2$ for each state.

It is interesting to note that even though Pennsylvania is the second largest natural gas producing state (after Texas) [17], the cost of natural gas (Figure 6) is relatively high compared to other states, and therefore, the cost of hydrogen produced via SMR is also high. As expected for natural gas producing states, Texas, Oklahoma, Louisiana, and West Virginia have low cost of natural gas and therefore low cost for hydrogen produced via SMR.

**Discussion**

The average LCOH is 50%-80% lower for SMR ($1/kg H2) and SMR+90%CCS ($1-2.5/kg H2) than electrolysis ($5/kg H2) in 2020. Even for the cheapest state for electrolysis in 2020, Oklahoma (~$3.5/kg H2), the cost is still higher than SMR+90% CCS. If the electricity cost in 2050 remains the same as 2020, LCOH of electrolysis ($2.5-3/kg H2) in 2050 is still higher than the LCOH of SMR+90% CCS in 2020. In 2050, almost 100% of LCOH of electrolysis is electricity cost. In order for electrolysis to break even with SMR+90% CCS, the costs of electricity need to decline to $0.02 /kWh. $0.02 is an achievable target since in 2021, the world has already seen low auction prices of USD 1.1 to 3 cents per kWh for renewables.

In 2020, the average carbon footprint of SMR (13 kg-CO2-eqv/kg-$H_2$) SMR+90% CCS (5 kg-CO2-eqv/kg-$H_2$) is lower than average carbon intensity of electrolysis (16-21 kg-CO2-eqv/kg-$H_2$). However, 5 states using SOEC have a lower carbon intensity than SMR+90% CCS in 2020 and 17 states of SOEC have a lower carbon intensity than SMR in 2020. If we adopt a linear decarbonisation of the grid until 2035, by 2025 the average carbon intensity of electrolysis will be lower than SMR and by 2030 the average carbon intensity of electrolysis will be lower than SMR+ 90% CCS.

Considering both average carbon intensity and average costs, SMR and SMR+90% CCS have lower costs of electrolysis and lower carbon intensity in 2020. However, in 2020 states such as Washington and Idaho with PEM would have a lower carbon footprint than SMR+90% CCS and twice the costs of SMR+90% CCS. However, if we assume off-peak electricity is only 50% cost of average rate and PEM electrolysis in both the states are only operating during off peak hours. The cost of PEM electrolysis will be lower than SMR+90% CCS.

Among all the three electrolysis technologies, SOEC has the highest costs and lowest carbon footprint now. It has the highest efficiency, while the manufacturing costs are high for pilot projects. However, if given the same production opportunity to drive down the cost curve, it will be the cheapest option in 2050. However, its advantage will diminish with lower electricity prices. With 2cent/kWh electricity, all three technologies will generate $1/kg H2 in 2050.

It is too early to pick the winning technology at the current stage. The Biden Administration's Infrastructure Bill allocates $8bn on Four Hydrogen Hubs (Figure 11). For Wyoming, Utah, Colorado and New Mexico ($4-5/kg H2; 30-45kg-CO2-eq/kg-$H_2$), the LCOH are relatively low while the carbon intensity is among the highest in the states. They could consider joint projects of renewables + electrolysis to reduce carbon intensity. For the Oklahoma, Arkansas, and Louisiana ($3-4/kg H2, 20 kg-CO2-eqv/kg-$H_2$) consortium, their costs and carbon intensity are lower than the four western states. Both Kentucky and West Virginia have low LCOH $5/kg H2. However, their carbon intensity is the highest among the states. They could consider SMR+CSS instead of electrolysis. For the four northeast states, New York is significantly better than Massachusetts, New Jersey, and Connecticut for electrolysis due to lowest costs and carbon intensity among the four. The two best states Washington and Idaho are missing in the plan, probably due to lack of demand. Due to high transmission costs, the selection of electrolysis plants should consider both production performances and end use demand.

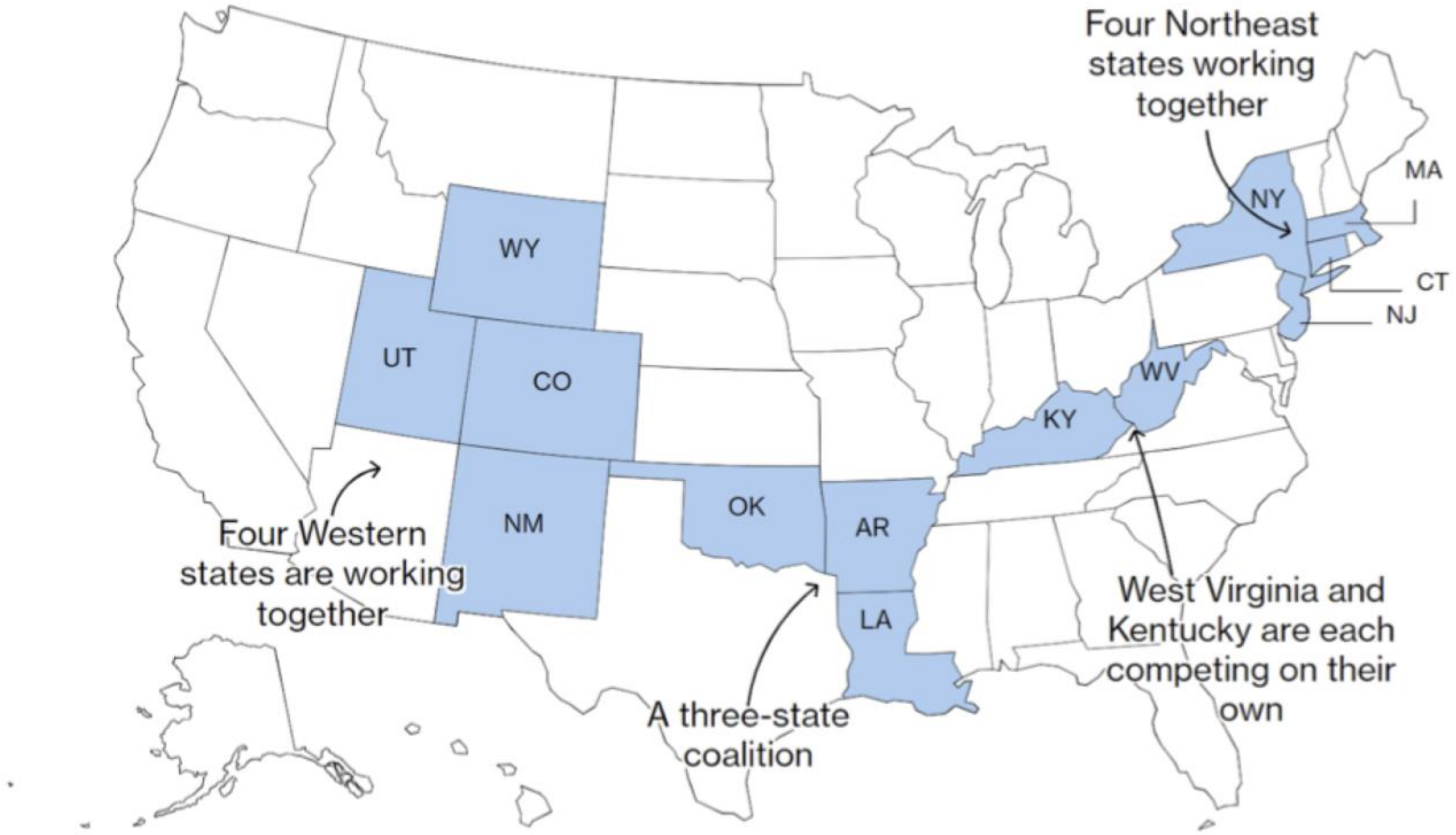

**Figure 12**. Hydrogen Hubs proposed in Infrastructure Bill [18].

**Conclusion**

Our results provide significant context to policy makers when deploying resources to spur hydrogen development. Our findings suggest that investing in electrolysis hydrogen production in Washington, Vermont, and Maine will result in low-cost hydrogen and low carbon intensity hydrogen.

Our findings demonstrate that in 2020 from an average cost and carbon intensity perspective, SMR with 90% CCS has a lower carbon intensity and LCOH than Solid Oxide Electrolyzer Cells (SOEC). However, this has the scope to improve in the future. Five states using SOEC have a lower carbon intensity than SMR with 90% CCS because they have a cleaner grid. As the US continues to invest in decarbonizing the grid, the cost and carbon intensity of electrolysis is expected to continue to fall. Electrolysis could be more competitive from a cost and carbon-intensity perspective by 2050. Assuming that the US is able to achieve 2 cent/kWh electricity and drive down the cost curve for electrolysis, hydrogen produced by any of the three electrolysis technologies can produce hydrogen at $1/kg, reaching parity with hydrogen produced through SMR with 90% CCUS.

Despite having a lower carbon intensity than SMR with 90% CCUS in some states, producing hydrogen through electrolysis faces significant headwinds. Even with the cost of hydrogen falling drastically to $1/kg

by 2050, hydrogen produced by electrolysis will still be at cost parity with hydrogen produced by SMR with 90% CCUS. Government intervention would be needed to make the production of lower carbon-intensity hydrogen more widespread.

Although it is too early to pick winners, the government should focus on setting stringent targets for cost and carbon-intensity and focus on demand-side policies. This will enable healthy competition between the different technologies and accelerate the learning curve.

Our state-by-state analysis of the levelized cost of hydrogen production by electrolysis provides important information that policy-makers can consider when developing a plan to scale hydrogen.

## Appendix

### Carbon Intensity of electrolysis

For the purposes of the paper, we chose to focus on carbon emissions emitted from using electricity to conduct electrolysis of water to produce hydrogen. The EPA E-grid Power Profiler tool was used to determine the carbon intensity of the electric grid of each state[19].

The total electricity used was multiplied by an emissions factor (kg CO2/kWh) to find the total amount of CO2 emitted. This was scaled by the amount of hydrogen produced over the lifecycle of the electrolyzer to find a measure for kg CO2/ kg H2 produced.